\documentclass{article}

\usepackage{romp34,bbm,amsmath,amsfonts,amssymb,latexsym}

\newcommand{\PCPsmwedge}{{\scriptstyle \wedge}}
\newcommand{\PCPqed}{\hspace*{\fill}\ensuremath{\Box}\\}

\title{A general construction of Poisson brackets on exact
  multisymplectic manifolds}
\author{ Michael Forger \\ 
  Departamento de Matem\'atica Aplicada,\\
  Instituto de Matem\'atica e Estat\'{\i}stica, \\
  Universidade de S\~ao Paulo, \\
  Caixa Postal 66281, \\
  BR--05311-970~ S\~ao Paulo, S.P., Brazil\\ 
  forger@ime.usp.br
  \\[2ex]
  Cornelius Pauf\/ler\hspace{3em} Hartmann R\"omer
  \\ 
  Physikalisches Institut\\
  Fakult\"at f\"ur Mathematik und Physik \\
  Albert-Ludwigs-Universit\"at Freiburg im Breisgau \\
  Hermann-Herder-Stra\ss e 3 \\
  D--79104~ Freiburg i.Br., Germany \\  
  cornelius.paufler@physik.uni-freiburg.de \hspace{3em} 
  hartmann.roemer@physik.uni-freiburg.de\\ 
}

\begin{document}

\maketitle
\begin{abstract}
In this note the long standing problem of the definition of a
Poisson bracket in the framework of a multisymplectic formulation of
classical field theory is solved. The new bracket operation can be applied to
forms of arbitary degree. Relevant examples are discussed and 
important properties are stated with proofs sketched.
\end{abstract}


\section{Introduction}

Multisymplectic geometry provides a mathematical framework to
describe classical field theory geometrically. Within this formulation
it is not necessary to break manifest Lorentz covariance nor  is there
a need to use concepts from infinite dimensional geometry. The
formalism dates back to the early work by De Donder, Dedecker, and
Weyl. By now the exploration of its geometrical aspects has reached
an elaborated stage, and a number of excellent reviews of this are
available to the reader (\cite{review}). In this article we propose an algebraic
structure that mimics closely the Poisson algebra of classical
mechanics.

Consider a classical field $\varphi^i(x)$, $i=1,\ldots,N$ with $N$
internal degrees of freedom over $n$-dimensional space-time $\mathcal
M$. Let $L(x^\mu,\varphi^i,\partial_\mu\varphi^i)$ denote the Lagrange
density. The corresponding equations of motion are ($\frac{d}{dx^\mu}$
denotes the total derivative w.r.t. $x^\mu$)
\begin{equation}
    \label{ELG}
 \frac{d}{d x^\mu} \,
 \left({\textstyle \frac{\partial L}{\partial \, \partial_\mu
       \varphi^i}}
 \right) \, - \,
 {\textstyle \frac{\partial L}{\partial \varphi^i}}~=~0.
\end{equation}
Introducing new field variables -- the {\sl multimomenta} -- by
\begin{equation}
    \label{PiMui}
    \pi^\mu_i={\textstyle \frac{\partial L}{\partial \partial_\mu \varphi^i}}
\end{equation}
and the {\sl De Donder-Weyl Hamiltonian} by
\begin{equation}
    \label{DWHam}
    H(x^\mu,\varphi^i,\pi^\mu_i)=\pi^\mu_i\,\partial_\mu\varphi^i-L
\end{equation}
one can easily verify that the equations of motion (\ref{ELG}) are
equivalent to the {\sl De Donder-Weyl equations} 
\begin{equation}
    \label{DWeq}
     {\textstyle \frac{\partial H}{\partial \pi_i^\mu}}~
 =~{\textstyle \frac {d}{d x^\mu}} \varphi^i \quad , \quad
 {\textstyle \frac{\partial H}{\partial \varphi^i}}~
 = \;- \, {\textstyle \frac{d}{d x^\mu}} \pi_i^\mu,
\end{equation}
provided certain regularity conditions hold for $L$. 

Note the remarkable similarity with the Hamiltonian equations of
motion of mechanics. The difference one encounters lies in the fact
that there are $n$ multimomenta associated to each degree of freedom
$\varphi^i$. Of course, for $n=1$ one recovers the case of
(time-dependent) Hamiltonian mechanics. 

Furthermore, as the fields are sections of some possibly non-trivial
fibre bundle $\mathcal E\rightarrow \mathcal{M}$, the notion of
derivatives $\partial_\mu\varphi^i$ of the fields does not exist in a
natural way. Rather, one has to use the theory of first jet bundles
(\cite{Saunders}) and -- later on -- their affine duals
$J^\star\mathcal{E}$.
For the sake of brevity we list the relevant objects occurring in
multisymplectic geometry in a table, thereby comparing the individual
items to their counterparts in the symplectic formulation of
time-dependent mechanics.\\
\begin{center}
\begin{tabular}{|r|l|}
\hline
time-dependent mechanics & field theory\\
\hline
extended configuration space&configuration bundle\\
$\mathcal{E}=\mathbbm R\times \mathcal{Q}$&
$\mathcal{E}\rightarrow\mathcal{M}^n$\\
\hline
extended tangent space&first jet bundle\\
$\mathbbm R\times T\mathcal{Q}$&$J^1\mathcal{E}$\\
\hline
$\mathcal{P}=(\mathbbm R\times T^\ast \mathcal{Q})\times \mathbbm R$
&
$P=J^\star \mathcal{E}$\\
$(t,q^i,p_i,E)$&$(x^\mu,q^i,p^\mu_i,p)$\\
\hline
canonical $1$-form &multicanonical form\\
$\theta=p_i\,dq^i+E\,dt$&
$\theta=p^\mu_i\,dq^i\PCPsmwedge d^nx_\mu+p\,d^nx$\\
\hline
symplectic form &multisymplectic form\\
$\omega=-d\theta= dq^i\PCPsmwedge dp_i- dE\PCPsmwedge dt$&
$\omega=-d\theta= dq^i\PCPsmwedge dp^\mu_i\PCPsmwedge d^nx_\mu
-dp\PCPsmwedge d^nx$\\
\hline
\end{tabular}
\end{center}
The multicanonical form $\theta$ can be defined in an intrinsic
way. We shall not elaborate on this but instead mention that there is a
vector field $\Sigma$ on $\mathcal{P}$, the scaling or Euler vector
field of the vector bundle $\mathcal{P}\rightarrow \mathcal{E}$, 
\begin{equation}
    \Sigma=p^\mu_i\,{\textstyle \frac{\partial}{\partial p^\mu_i}}
    +p\,{\textstyle \frac{\partial}{\partial p}},
\end{equation}
that
satisfies 
\begin{equation}
    i(\Sigma)\omega=-\theta.
\end{equation}
$\mathcal{P}$, together with $\omega$ and $\theta$, is an example of
an {\sl exact multisymplectic manifold}. Note in particular that
$\omega$ is non-degenerate on vector fields.

There is a natural projection from the extended multiphase space
$\mathcal{P}$ to the {\sl ordinary multiphase space}
$\tilde{\mathcal{P}}$ given by (we just write down the coordinate expression)
\begin{equation}
        \mathcal{P}\rightarrow \tilde{\mathcal{P}},\quad
        (x^\mu,q^i,p^\mu_i,p)\mapsto (x^\mu,q^i,p^\mu_i).
\end{equation}
The dimension of $\tilde{\mathcal{P}}$ is one less than that of
$\mathcal{P}$. The ordinary multiphase space has been used by
Kanatchikov (\cite{Kanatchikov}), Echeverr\'{\i}a-Enr\'{\i}quez et al.
(\cite{EE}) in their work. 
It shows the unpleasant feature
of not carrying a canonical multisymplectic structure. However, with
the help of connections in $\mathcal{E}$ and $T\mathcal{M}$, the
analogues $\theta^V$ and $\omega^V$ of the above items can be defined
and one has 
\begin{equation}
    \omega^V=-d^V\theta^V,
\end{equation}
where $d^V$ is an exterior derivative along the fibres of
$\tilde{\mathcal{P}}\rightarrow\mathcal{M}$ acting on forms on
$\tilde{\mathcal{P}}$.

\section{Poisson forms}

It is natural to study solutions of the equation
\begin{equation}\label{iXomega}
    i(X_f)\,\omega=df.
\end{equation}
As the multisymplectic form $\omega$ is an $(n+1)$-form, $f$ can be an
$r$-form, $r=0,1,\ldots,(n-1)$. Correspondingly, $X_f$ has to be an
$(n-r)$-vector field, $X_f\in\Lambda^{n-r}\mathfrak
X(\mathcal{P})$. Note that $\omega$ is degenerate on multi-vector
fields of tensor degree higher than $1$ (otherwise it would have to be
the volume form on $\mathcal{P}$). Consequently, $X_f$ is not uniquely
determined by $f$, nor is $f$ fixed by $X_f$. Moreover, it is known
that there are conditions on $f$ which have to be met for an $X_f$ to
exist. These restrictions concern the dependence of the coefficients
of $F$ on the multimomenta $p^\mu_i$. If the pair $(X_f,f)$ forms a
solution of (\ref{iXomega}) then $f$ is called {\sl Hamiltonian form}
and $X_f$ {\sl Hamiltonian multi-vector field}. With the definition of
the Lie derivative along an $r$-vector field $Y$,
\begin{equation}
    L_Y=d\,i(Y)-(-1)^ri(Y)\,d
\end{equation}
one can formulate necessary and sufficient conditions for an
multi-vector field $X$ to be a Hamiltonian multi-vector field
associated to some Hamiltonian form. More precisely, if $X$ denotes such
a Hamiltonian multi-vector field, then 
\begin{equation}
    L_X\omega=0,
\end{equation}
and if 
\begin{equation}\label{exact}
    L_X\theta=0
\end{equation}
then $X$ is a Hamiltonian r-vector field associated to 
\begin{equation}
    f=J(X)=(-1)^ri(X)\,\theta.
\end{equation}
Multi-vector fields satisfying (\ref{exact}) are called {\sl exact
  Hamiltonian multi-vector fields}.\\
 Typical examples of Hamiltonian
forms will be discussed below.

As already mentioned the Hamiltonian multi-vector field $X_f$ is not
fixed by $f$. However, as we shall see in a moment, the definition of
a bracket operation between 
Hamiltonian forms will involve contractions of Hamiltonian
multi-vector fields with Hamiltonian forms. This results in an
undesirable dependence on the choice of the Hamiltonian multi-vector
field. We therefore restrict the Hamiltonian forms further.
\begin{definition}{Definition}
    A {\sl Poisson form} $f$ on $\mathcal{P}$ is a Hamiltonian form
    that in addition 
    satisfies 
    \begin{equation}
        i(Y)\,\omega=0\quad\Rightarrow\quad i(Y)\,f=0
    \end{equation}
   for all multi-vector fields $Y$ on $\mathcal{P}$. Equivalently,
   for a Poisson form $f$ 
   there exists a multi-vector field $Z$ on $\mathcal{P}$ with
   \begin{equation}
       i(Z)\,\omega=f.
   \end{equation}
\end{definition}
{\bf Remark.} For exact Hamiltonian $r$-vector fields $X_f$, one has
\begin{equation}
    J(X_f)=i(X_f\PCPsmwedge \Sigma)\,\omega.
\end{equation}

To show that the notion of a Poisson form is non-empty we give a list
of examples that have been discussed in the literature.

Functions on $\mathcal{P}$ are Poisson. If in addition a function $f$
is of the form
\begin{equation}
    f(x^\mu,q^i,p^\mu_i,p)=-H(x^\mu,q^i,p^\mu_i)-p,
\end{equation}
then it admits a Hamiltonian $n$-vector field $X$ that is locally
decomposable, 
\begin{equation}
    X=Z_1\PCPsmwedge\cdots\PCPsmwedge Z_n, \quad Z_\mu\in\mathfrak X(\mathcal{P}).
\end{equation}
The solutions of the De Donder-Weyl equations with $H$ as above define
a tangent space at every point of $\mathcal{P}$ that is spanned by the
collection of the $Z_\mu$. The converse, i.e. the integration of the
$Z_\mu$ to a solution of the De Donder-Weyl equations can be performed
if additional conditions are fulfilled (for details, see \cite{PR}). 

A form $\tilde f$ on the ordinary multiphase space
$\tilde{\mathcal{P}}$ is called horizontal if it vanishes on vector
fields that are vertical w.r.t. the projection onto
$\mathcal{M}$. Kanatchikov
has studied such horizontal forms on $\tilde{\mathcal{P}}$ that in
addition satisfy
\begin{equation}\label{HamMvfKan}
    i(\tilde{X}_{\tilde f})\,\omega^V=d^V\tilde f.
\end{equation}
In the sequel we will refer to such forms as {\sl Hamiltonian forms 
\`a la Kanatchikov}.
One can show that they are in $1-1$ correspondence with
horizontal Hamiltonian forms $f$ on $\mathcal{P}$, where $f$ is the
pullback of $\tilde f$. Obviously, horizontal Hamiltonian forms on
$\mathcal{P}$ are Poisson.

Let $\xi_{\mathcal{E}}$ be a vector field on the configuration bundle 
$\mathcal{E}$ which is projectable onto $\mathcal{M}$. There is a
canonical lift of $\xi_{\mathcal{E}}$ to $\mathcal{P}$. The resulting
vector field $\xi_{\mathcal{P}}$ is the generator of a special
canonical transformation on $\mathcal{P}$. One has
\begin{equation}
    L_{\xi_{\mathcal{P}}}\,\theta=0.
\end{equation}
Consequently,
\begin{equation}
    J(\xi_{\mathcal{P}})=i(\xi_{\mathcal{P}})\,\theta
\end{equation}
is Poisson. Moreover, let $X_1$, $X_2$ be {\sl commuting} 
exact Hamiltonian vector fields. Then one can show
\begin{equation}
    L_{X_1\PCPsmwedge X_2}\,\theta=0,
\end{equation}
and one obtains examples $J(X_1\PCPsmwedge X_2)$ of Poisson forms of
intermediate tensor degree. These forms can be viewed as the
counterparts to higher dimensional orbits of transformations.

There is a canonical projection from Hamiltonian $(n-r)$-forms onto Poisson
$(n-r)$-forms with exact Hamiltonian $r$-vector fields, given by
\begin{equation}
        f\mapsto (-1)^{r-1}i(X_f)\,\theta,
        \quad 
        X_f\mapsto X_f+[\Sigma,X_f].
\end{equation}
Obviously, this map does have a kernel. However, it has not been
clarified yet whether there exist Hamiltonian 
multi-vector fields that are not Poisson.

\section{Poisson brackets}

Before we turn to the definition of a bracket operation between Poisson
forms let us briefly recall the canonical extension of the Lie-bracket
of vector fields to the algebra of multi-vector fields. This
composition is known under the name Schouten bracket. The defining
formul\ae\  are (let $X,Y,Z$ be multi-vector fields of respective tensor degree
$r,s,t$)
\begin{equation}
    \begin{split}
         [X,Y \PCPsmwedge\, Z\>\!]&=[X,Y] \;\PCPsmwedge\; Z \, + \,
         (-1)^{(r-1)s} \, Y \>\!\PCPsmwedge\;
         [X,Z\>\!],
         \\
         [\>\!Y,X]&=- \, (-1)^{(r-1)(s-1)} \, [X,Y].
     \end{split}
\end{equation}
Moreover, one requires that it coincide with the standard
Lie bracket on vector fields. This fixes the bracket. An important
consequence of these properties is a graded Jacobi identity
\begin{equation}\label{Jacobi-mvf}
    (-1)^{(r-1)(t-1)} \, [X,[Y,Z]] \; + \; \mathrm{cyclic~perm.}~=~0~.
\end{equation}
In addition, one can derive the following formula relating Lie
derivative and Schouten bracket:
\begin{equation}\label{useful}
    L_{[X,Y]}
    ~=~(-1)^{(r-1)(s-1)} \, L_X^{} L_Y^{}  - \,
    L_Y^{} L_X^{} .
\end{equation}

Now lets turn to the definition of a bracket operation for Poisson
forms. 
Let $f$, $g$, $h$ be Poisson forms with associated Hamiltonian
multi-vector fields $X_f$, $X_g$, and $X_h$ of respective tensor
degree
$r$, $s$, $t$.\\
An obvious ansatz for the Poisson bracket is given by
\begin{equation}\label{bracket1}
    \{f,g\}'=(-1)^{r}i(X_f)\,i(X_g)\,\omega.
 \end{equation}
This formula has been suggested by many authors. It shows the desired
properties 
\begin{align}\label{antisymmetry}
    \{f,g\}'&=-(-1)^{(r-1)(s-1)}\{g,f\}'\\
    \label{MvfOfBracket}
    X_{\{f,g\}'}&=[X_g,X_f].
\end{align}
However, it does not satisfy a graded Jacobi identity that meets with
(\ref{Jacobi-mvf}):
\begin{equation}\label{Jacobi1}
    \begin{split}
        (-1)^{(r-1)(t-1)}\{f,\{g,h\}'\}'+\textrm{cycl. perm.}
        &
        =(-1)^{(t-1)(r-1)+s}
        d(i(X_f)\, i(X_g)\, i(X_h)\, \omega).
    \end{split}
\end{equation}
Moreover, hideous additional terms occur in the composition of two
generators of transformations (let $X$, $Y$ be exact Hamiltonian
multi-vector fields of degree $r$ and $s$),
\begin{equation}\label{equivariance}
    \{J(X),J(Y)\}'=J([Y,X])-(-1)^{(s-1)r}d(i(Y)\,i(X)\,\theta.
\end{equation}
If $d$ has trivial cohomology one can use (\ref{Jacobi1}) as a
starting point for the construction of an $l(\infty)-$structure. 

We propose a modification of (\ref{bracket1}) which preserves
(\ref{antisymmetry}) and (\ref{MvfOfBracket}) 
but cures
the anomalous terms in (\ref{Jacobi1}) and (\ref{equivariance}).
\begin{definition}
    {Definition}
    Let $f$ and $g$ be Poisson forms with respective Hamiltonian
    multi-vector fields $X_f$ and $X_g$. Then their bracket $\{f,g\}$
    is given by
 \begin{equation} \label{eq:POISB1}
     \{f,g\}~=~- \, L_{X_f} g \, + \, (-1)^{(r-1)(s-1)} L_{X_g} f \,
            - \, (-1)^{(r-1)s} L_{X_f \wedge\>\! X_g} \, \theta~,
 \end{equation}
 Where $X_f$ and $X_g$ are multi-vector fields of degree $r$ and $s$,
 respectively. 
 \\
 Equivalently,
 \begin{equation} \label{eq:POISB2}
     \begin{split}
         \{f,g\} =&
         (-1)^{r} \, i(X_g)\, i(X_f)\,\omega\\
         &+ \; d \,
         \Bigl( (-1)^{(r-1)(s-1)} \, i(X_g)\, f \, - \, i(X_f)\,
         g \, 
         - \,(-1)^{(r-1)s} \, i(X_g)\, i(X_f)\, \theta \, \Bigr).
     \end{split}
 \end{equation}
\end{definition}
\begin{theorem}
    {Proposition} Let $f$, $g$, $h$ be Poisson forms with respective
    Hamiltonian 
    $r$-, $s$-, and $t$-vector fields $X_f$, $X_g$ and $X_h$. Then the following properties hold.
    \begin{enumerate}
    \item The bracket $\{f,g\}$ does not depend on the particular
        choice of $X_f$ and $X_g$.
    \item $X_{\{f,g\}}=[X_g,X_f]$.
    \item The bracket $\{f,g\}$ of two Poisson forms $f$, $g$ is again a
        Poisson form.
    \item The bracket is graded antisymmetric,
        \[\{f,g\}=-(-1)^{(r-1)(s-1)}\{g,f\}.\] 
    \item It satisfies a graded Jacobi identity,
        \[
        (-1)^{(r-1)(t-1)} \{f,\{g,h\}\} \; + \;
        \mathrm{cyclic~perm.}~=~0~.
        \]
    \item For exact Hamiltonian multi-vector fields $X$, $Y$, one has 
        \[
        \{J(X),J(Y)\}=J([Y,X]).
        \]
    \item Let $\tilde f$, $\tilde g$ be Hamiltonian forms \`a la
        Kanatchikov, $\tilde X_{\tilde f}$ and $\tilde X_{\tilde g}$
        be their multi-vector fields (of tensor degree $r$ and $s$) 
        according to (\ref{HamMvfKan}). 
        Then $i(\tilde X_{\tilde f})\,i(\tilde X_{\tilde
          g})\,\omega^V$
        is again Hamiltonian \`a la Kanatchikov and can be pulled back
        to $\mathcal{P}$, where it coincides with the bracket
        $\{f,g\}$ of the pulled back forms $f$ and $g$.
    \end{enumerate}
\end{theorem}
{\bf Proof.} While the proofs of items 1,2,3,4,6 are obvious, the
demonstration of the graded Jacobi identity is rather lengthy. It
follows from the properties (\ref{Jacobi-mvf}) and (\ref{useful}). A
calculation in coordinates shows the last statement. The details of
the proofs are contained in \cite{FPR}.\PCPqed
{\bf Remark.} The definition and the proposition (except the last
statement)
do not involve any properties other than 
\begin{equation}
\omega=-d\theta,\quad i(\Sigma)\,\omega=-\theta.
\end{equation}
Hence the construction can be carried over to arbitrary exact
multisymplectic manifolds, in particular to the direct treatment of
higher order Lagrangeans (\cite{Gotay}).\\
The notion of a Poisson bracket is not completely justified as no
product structure has been identified yet. One can show that a
horizontal Poisson form of degree $r$ has to be polynomial in the
multimomenta of degree $(n-r)$ or less. Hence, the product of a
Poisson $r$-form and a Poisson $s$-form can be at most an
$(r+s-n)$-form. There are two candidates for such a composition:
Kanatchikov's $\bullet$-product
\begin{equation}
    f\bullet g=\ast^{-1}(\ast f\PCPsmwedge \ast g),
\end{equation}
where $\ast$ is a Hodge operation on $\mathcal{M}$, acting on
horizontal forms on $\tilde{\mathcal{P}}$, and
-- on the side of multi-vector fields -- the map
\begin{equation}
    X,Y\mapsto X\PCPsmwedge Y.
\end{equation}
One can show that both maps do not coincide. While the former does
yield a Poisson form and satisfies a graded Leibniz rule, the latter
cannot be carried over to Poisson forms. This can be seen from the identity
\begin{equation}
    L_{X_f\wedge X_g}\omega=\pm d\{f,g\}.
\end{equation}
The $\bullet$-product, however, cannot be extended to general Poisson
forms without the use of a connection. Moreover, the product of two
functions will always give zero which makes it impossible to view it
as a generalisation of the product of functions in symplectic geometry.

\section{Conclusions}

Multisymplectic geometry provides a covariant formalism for classical
field theory. It uses elements from finite dimensional differential
geometry throughout. There is a close resemblance to the formulation
of time dependent Hamiltonian mechanics. However, there are a number
of features that do not occur in mechanics. 

In particular there is a correspondence between forms and multi-vector
fields that replaces the interplay of functions and Hamiltonian vector
fields of symplectic geometry. Forms of degree $n-1$, where $n$ is the
dimension of space-time, come in very naturally when considering
symmetry transformations. Remember that according to Noether's theorem
there is a conserved charge for every symmetry, and this charge is
obtained through integration over $(n-1)$-dimensional hypersurfaces of
space-time. On the other hand, as solutions of the field equations are
objects of dimension higher than $1$, we expect $n$-vector fields to
replace the Hamiltonian vector fields of mechanics. This is indeed the
case. 

These considerations show that there is a need to have a (Poisson)
bracket operation for forms at hand, 
and in view of later quantisation attempts
this structure should obey a Jacobi identity. In this note we have
proposed such a bracket for Poisson forms. We have shown that all
relevant examples occurring in the geometrical formulation of classical
field theories are indeed Poisson forms. Moreover, the proposed
operation does satisfy a graded Jacobi identity that is closely
related to the Jacobi identity of the Schouten bracket of multi-vector
fields. We have shown that the bracket operation closes on Poisson
forms and contains as a particular case the Gerstenhaber algebra
introduced by Kanatchikov (\cite{Kanatchikov}). Furthermore, with this
structure at hand, one is now able to algebraically investigate
symmetry transformations, including space-time translations.

The construction is independent of the particular structure of the
exact multisymplectic manifold under consideration. Therefore, it is
possible to apply the results to field theories beyond first order
Lagrangeans, i.e. to gravity as an example. It would be of interest to
study the treatment of constraints in this context. These matters are
currently under investigation.

\end{document}